\documentclass[12pt]{article}

\usepackage{amsmath,amsfonts,bbold}
\usepackage{latexsym}
\usepackage{amssymb}
\usepackage{epsfig,psfrag,cite}
\usepackage{stmaryrd}

\makeatletter   
\@addtoreset{equation}{section}   
\makeatother

\newcommand{\be}{\begin{equation}}  
\newcommand{\ee}{\end{equation}}  
\newcommand{\bea}{\begin{eqnarray}}  
\newcommand{\eea}{\end{eqnarray}}  
\newcommand{\Tr}{\operatorname{Tr}}
\newcommand{\diag}{\operatorname{diag}}

\newcommand{\nn}{\nonumber}

\newcommand{\one}{\mathbb{1}}

\begin{document}

\thispagestyle{empty}
\vspace*{.5cm}
\noindent

\vspace*{1.9cm}

\begin{center}
{\Large\bf A New Class of Rank Breaking Orbifolds}
\\[2.3cm]
{\large G.~von Gersdorff}\\[.5cm]
{\it Department of Physics and Astronomy, Johns Hopkins University, 
3400 N Charles Street, Baltimore, MD 21218}
\\[.4cm]
{\small\tt (\,gero@pha.jhu.edu\,)}
\\[1.3cm]

{\bf Abstract}\end{center} We describe field-theory $T^2/Z_n$
orbifolds that offer new ways of breaking $SU(N)$ to lower rank
subgroups.  We introduce a novel way of embedding the point group into
the gauge group, beyond the usual mapping of torus and root lattices.
For this mechanism to work the torus Wilson lines must carry
nontrivial 't Hooft flux.  The rank lowering mechanism proceeds by
inner automorphisms but is not related to continous Wilson lines and
does not give rise to any associated moduli.  We give a complete
classification of all possible $SU(N)$ breaking patterns.  We also
show that the case of general gauge group can already be understood
entirely in terms of the $SU(N)$ case and the knowledge of standard
orbifold constructions with vanishing 't Hooft flux.

\newpage
\section{Introduction}

Orbifolds \cite{Dixon:1985jw} are one of the most explored avenues in
the study of string theory compactifications. Not only do they possess
phenomenologically appealing features such as chirality, reduced
supersymmetry, and a built-in gauge symmetry beaking mechanism, they
are also extremely tractable and provide a welcome starting
point to study more complicated vacua though string theory's many
dualities.  Notwithstanding, the classification of all orbifold vacua
of the (heterotic) string seems to be an extremely difficult task, and
the search of the standard model, or its supersymmetric extension, in
this vast ``landscape'' of vacua has only been partially successful.

A more modest approach, justified in its own right, are orbifold grand
unified theories (orbifold GUTs).  It is quite conceivable that some
of the extra dimensions are larger than others, and intermediate models
with effectively fewer extra dimensions could be realized in
nature. In view of this, a lot of effort has been made to construct
five and six dimensional models that break the GUT group by orbifolding
down to the SM
\cite{Kawamura:2000ev,
Asaka:2002nd,
Hebecker:2003jt,
vonGersdorff:2006nt
}.
Some intermediate 6d models appearing as particular compactification
limits of the heterotic string have been described in
Ref.~\cite{Buchmuller:2005jr}.

A challenge in obtaining the standard model gauge group by orbifolding
is the fact that the simplest consistent choices for the twists do not
reduce the rank of the gauge group. In heterotic string theory, the
anomaly-free gauge groups have rank 16 while the Standard Model only
has rank 4. Rank reduction usually proceeds through one of the
following mechanisms
\begin{itemize}
\item
Continous Wilson lines \cite{Ibanez:1987xa,Forste:2005rs}: A given
orbifold vacuum can possess a nontrivial moduli space in the gauge
sector, i.e.~flat directions in the tree level potential for the
extra dimensional components ($A_{4,5\dots}$) of the gauge bosons. The
latter typically transform in non-adjoint representations of the gauge
group left unbroken by the orbifolding. By obtaining vacuum
expectation values they can break the gauge symmetries further,
thereby reducing its rank. From a four dimensional (4d) point of view,
this is nothing but the standard Higgs mechanism. This idea has been
applied in the context of electroweak symmetry breaking and is often
referred to as ``gauge-Higgs unification''
\cite{Manton:1979kb,Hosotani:1983xw,Krasnikov:1991dt,Antoniadis:2001cv,Kubo:2001zc,vonGersdorff:2002as
,Csaki:2002ur,Burdman,Biggio:2004kr,Martinelli:2005ix}.  The flatness
is lifted at loop level by a finite and calculable potential
\cite{Hosotani:1983xw}, giving rise to a discrete set of
vacua. Unfortunately, in many circumstances, the vacuum calculated
this way actually corresponds to a particular point in moduli space
where the rank of the gauge group is restored
\cite{Kubo:2001zc,vonGersdorff:2002as }. Moreover, some Higgs mass
terms localized at the fixed point are unprotected by the surviving
gauge symmetry \cite{Csaki:2002ur,Biggio:2004kr} and can destroy the
finiteness and predictivity of the model.

\item
Green-Schwarz mechanism: If the unbroken gauge group contains
anomalous $U(1)$ factors, the latter can be spontaneously broken by an
orbifold version~\cite{Horava:1995qa} of the Green-Schwarz mechanism
\cite{Green:1984sg}. This mechanism is realized, e.g.~, in the model
of Ref.~\cite{Antoniadis:2001cv}, where the rank-6 group $U(3)^2$ was
broken to the Standard model by the presence of two anomalous $U(1)$
symmetries.

\item
Additional Higgs multiplets at the fixed points, as, e.g., in
Ref.~\cite{Asaka:2002nd}.

\item
Outer automorphisms. A particular choice of the gauge twists,
corresponding to a symmetry of the Dynkin diagram of the associated
Lie algebra, can break the rank. There are only finitely many
possibilities.

\end{itemize}

In this paper we want to introduce a new way to break the rank of the
gauge group by orbifolding.  We will mainly restrict ourselves to
$T^2/Z_n$ orbifolds with gauge group $SU(N)$ and will comment on
generalizations to higher dimensional tori and other gauge groups in
Sec.~\ref{disc}.  An orbifold is specified by the gauge twists
associated to translations and rotations of the underlying torus
lattice. The spacetime translations commute, and so must the
corresponding twists. However, in a pure gauge theory, the fields
transform in the adjoint representation, and the twists need only
commute up to an element of the center of the group. This yields
nontrivial gauge bundels on the torus which still have a flat gauge
connection (i.e.~the corresponding field strength vanishes)
\cite{'tHooft:1979uj}. The center of $SU(N)$ is isomorphic to $Z_N$.
Hence, there are $N$ physically different disconnected vacua, or, more
precisely, the moduli space consists of $N$ disconnected componenents.
The nontrivial statement we make in this paper is that one can
orbifold these configurations.  Since the distinction to the standard
orbifold construction is quite essential, let us dwell a little more
on this point. In the standard approach, lattice translations are
realized by shift vectors, i.e.~the corresponding holonomies exactly
commute and can be realized as elements of the same Cartan torus. The
rotations of the torus lattice are then realized by an element of the
Weyl group (rotations of the root lattice). Here, instead, the lattice
translations are already realized as rotations of the root lattice, in
a way that makes it impossible to choose a Cartan torus such that both
of them simultaneously become shifts. Consequently, the orbifold
twists associated to the rotations of the torus lattice cannot be
related to any symmetry of the root lattice used to define the torus
holonomies.

The paper is organized as follows. In Sec.~\ref{torons} we review the
nontrivial flat $SU(N)$ gauge bundles on the two-torus, give an
explicit form for the holonomies, and describe their symmetry breaking
patterns.  We also explain how other gauge groups can be treated once
the $SU(N)$ case is known.  These gauge bundles are orbifolded in
Sec.~\ref{orbifold}. In Sec.~\ref{m0} we treat first the case
$m=0$. This does not involve any new concepts, but we include it here
for completeness and comparison. Also, in App.~\ref{moduli} we compute
the moduli space for this case. In Sec.~\ref{generic} we calculate the
orbifold twists for the generic case, making use of the results
obtained in Sec.~\ref{torons} and Sec.~\ref{m0}. Finally, in
Sec.~\ref{disc} we summarize our results and discuss some
applications.

\section{Breaking $SU(N)$ on $T^2$: Torons.}
\label{torons}

In this section we would like to recall 't Hooft's toron
configurations \cite{'tHooft:1979uj}.  These are simply flat $SU(N)$
gauge bundles on the torus, which can be characterized by their
holonomies. Upon shifts in the torus lattice
\be
z\to z+\lambda
\ee
gauge fields are identified up to gauge transformations \footnote{We
make use of the fact that we can choose a gauge where the transition
functions are $z$-independent, see, e.g., Ref.~\cite{Salvatori:2006pb}.}
\be
A_M(z+\lambda)=T_\lambda A_M(z) T_\lambda^{-1}\,.
\label{torus}
\ee
It is clearly sufficient to restrict to the two lattice-defining base
vectors $\lambda_{1,2}$. As lattice translations commute, the commutator
of the two transition functions has to act as the identity.
\be
T_1 T_2 T_1^{-1}T_2^{-1}=e^{2\pi i \frac{m}{N}}\,.
\label{hooft}
\ee
On the right hand side we have allowed for a general element of the center
of the group, which, for $SU(N)$, equals $Z_N$. Such a gauge
transformation indeed acts trivially on the adjoint representation the
gauge fields transform in.  The integer quantity $m$ is called the 't
Hooft nonabelian flux.  We stress that it is in principle possible to
simultaneously diagonalize the matrices $T_1$ and $T_2$ in the
adjoint.\footnote{For an explicit diagonal basis see
Ref.~\cite{Salvatori:2006pb}.}  For nonzero $m$, it is not possible to
represent both $T_i$ as elements of the same Cartan torus.  It is,
however, possible to choose a Cartan torus left fixed (though not
pointwise fixed) by both $T_i$. As a consequence, one can
realize the $T_i$ as Weyl group elements w.r.t.~the same Cartan
subalgebra.

The flux $m$ (more precisely the phase appearing on the r.h.s.~in
Eq.~(\ref{hooft})) labels the equivalence classes of the transition
functions and determines the vacua of the theory.  We would like to
find the unbroken subgroup for each vaccum, i.e.~we are looking for
the generators that are left invariant by the action of the $T_i$:
\be
T_i \mathcal T T_i^\dagger=\mathcal T\,.
\ee
For fixed $m$, there is still a continous degree of freedom in
choosing the $T_i$, even within the gauge where the transition
functions are constants: If, for a particular solution to
Eq.~(\ref{hooft}), the unbroken subgroup $\mathcal H$ is nontrivial,
one can always turn on Wilson lines in the Cartan torus of $\mathcal
H$ and still obtain a solution with the same value for $m$. Such an
additional Wilson line will lead to a different subgroup $\mathcal
H'$, however, the rank of $\mathcal H$ and $\mathcal H'$ must remain
the same.  This freedom is related to the fact that each vacuum will
in general possess a nonzero moduli space, i.e. flat directions in
the potential for the extra dimensional components of $A$.

To describe the solutions, one decomposes $N$ and $m$ according to
their greatest common divisor $K={\rm g.c.d}(N,m)$. Explicit solutions to
Eq.~(\ref{hooft}) are then given by \cite{Salvatori:2006pb,vanBaal:1983eq}
\be
T_1=Q_{N/K}\otimes \one_{K}\,,\qquad (Q_{L})_{jk}=q_L^{-(L-1)/2}\delta_{j,k-1}\,,
\label{T1}
\ee
\be
T_2=(R_{{N/K}})^{ m/K}\!\otimes \one_{K}\,,\qquad (R_L)_{jk}= q_L^{-(L-1)/2+j-1}\delta_{j,k}\,,
\label{T2}
\ee
where $q_L=\exp(2\pi i/L)$. The index on $Q$, $R$ and $\one$
indicates the dimensionality of the matrices and the Kronecker
$\delta$ is assumed to be periodic.  The matrices $Q$ and $R$ satisfy
\be
Q\,R=q\,R\, Q\,,  \qquad Q^{L}=R^{L}=(-)^{{L}-1}\one\,.
\ee
Hence, Eq.~(\ref{T1}) and Eq.~(\ref{T2}) are a particular
solution to Eq.~(\ref{hooft}). It can then be shown
that the twists $Q_{N/K}$ and $(R_{N/K})^{m/K}$ break $SU({N/K})$ completely
\cite{Salvatori:2006pb}.  Writing the generators of $SU(N)$ as
\be
\mathcal T_N\in\{\mathcal T_{N/K}\otimes \mathcal T_K,\ 
\one_{N/K}\otimes \mathcal T_K,\ \mathcal T_{N/K}\otimes \one_K\}\,.
\ee
We immediately read off that the unbroken subgroup is generated by
$\one_{N/K}\otimes\mathcal T_K$ and, thus, is $SU(K)$.  The most
general solution to Eq.~(\ref{hooft}) can then be obtained by
replacing the unit matrices in Eq.~(\ref{T1}) and (\ref{T2}) with
commuting Wilson lines of $SU(K)$, which one can take to be elements
of the same Cartan torus:
\begin{eqnarray}
T_1&=&Q_{N/K}\otimes \exp(2\pi i\, W_1)\,,
\nn
\\
T_2&=&(R_{N/K})^{m/K}\!\otimes \exp(2 \pi i\, W_2)\,.
\label{wilson}
\end{eqnarray}
The shift vectors $W_1$ and $W_2$ are elements of the Cartan
subalgebra of $SU(K)$. Nontrivial $SU(K)$ Wilson lines further break
$SU(K)$, but do not reduce its rank. In summary, a toron configuration
with $SU(N)$ flux $m$ can be decomposed into a toron configuration
with $SU(N/K)$ flux $m/K$ and an $SU(K)$ configuration with
vanishing flux.

However, we would like to stress here that different $SU(K)$ Wilson
lines, strictly speaking, do not correspond to different physical
theories. The reason is that the above mentioned flat directions are
lifted at the quantum level and two such theories will dynamically
evolve to the same vacuum. One can always perform a field
redefinition, corresponding to a nonperiodic gauge transformation that
removes the continous Wilson line but generates a vacuum expectation
value (VEV)
\begin{align}
A_4&=W_1\,,\nn\\
A_5&=W_2\,.
\label{hosotani}
\end{align}
One sees that such a field redefinition induces a shift along a flat
direction. In other words, a theory with nonzero Wilson line and a
given point in the moduli space is equivalent to a vanishing Wilson
line and a shifted point in moduli space. The degeneracy of the flat
directions is lifted at the quantum level. The effective potential
clearly only depends on the sum of the Wilson line induced background,
Eq.~(\ref{hosotani}), and the explicit background, and the true vacuum
of two theories with different continous Wilson lines coincide.
It is important to realize that there is no analogous field
redefinition that could change the value of $m$:\footnote{By enforcing
such a field redefinition to, say, remove the Wilson line $T_1$, the
other transition function would no longer remain constant.} Two vacua
with different $m$ are truly disconnected.

The natural question to ask is whether all this can be generalized to
gauge groups other than $SU(N)$. This question has been extensively
discussed in Ref.~\cite{Borel:1999bx}, see also
Refs.~\cite{Schweigert:1996tg,Lerche:1997rr,Kac:1999gw}.  Here we only
give some heuristic arguments and some examples.  A necessary and
sufficient condition for the existence of nontrivial 't Hooft flux is
that the group possesses nontrivial center.\footnote{More precisely,
the center of the universal cover, which is isomorphic to the
fundamental group of the adjoint representation.}  This is true for
the $SO(N)$ and $Sp(2N)$ groups, as well as for the exceptional groups
$E_6$ and $E_7$. The center can always be embedded in suitable $SU(N)$
subgroups \cite{Borel:1999bx}, and, hence, the above construction can
be carried out straightforwardly.  Trivial examples are the groups
$SO(3)$, $SO(4)$, $SO(6)$, and $Sp(2)$ that are actually isomorphic to
some special unitary groups.  For a nontrivial example take $SO(8)$
whose center is $C=Z_2\times Z_2$. Consider now the maximal subgroup
$SU(2)^4\subset SO(8)$. By inspection of the branching rules for the
$SO(8)$ irreducible representations ${\bf 8_v}$ and {$\bf 8_s$}, one
can see that a suitable parametrization of the two $Z_2$'s of the
center is
\be
c_1=(-\one,-\one,\one,\one)\,,\qquad c_2=(\one,-\one,-\one,\one)\,,
\ee
where $\pm \one$ represent the center of the corresponding $SU(2)$
factor. The branching of the adjoint is
\be
{\bf 28}\to ({\bf 3,1,1,1})+({\bf 1,3,1,1})+({\bf 1,1,3,1})
+({\bf 1,1,1,3})+({\bf 2,2, 2,2})\,.
\ee
It can be directly verified that $C$ acts trivially on the {\bf 28},
as it must.  For a given $c\in C$, particular solutions for $T_1(c)$
and $T_2(c)$ can now be constructed by making use of the results for
$SU(2)$. While any pair $T_i(c)$ clearly projects out two of the four
triplets, the action on the fourfold doublet requires a more careful
analysis. Take, for instance $c=c_1$, then the standard solution acts
on the ({\bf 2,2,2,2}) as
\be
T_1=\sigma_1\otimes\sigma_1\otimes\one\otimes\one,\qquad 
T_2=\sigma_3\otimes\sigma_3\otimes\one\otimes\one\,.
\ee
The two twists can be diagonalized simultaneously. There are four
eigenstates, each transforming as ({\bf 2,2}) of the surviving
$SU(2)^2$. One of these eigenstates has unit eigenvalue on both $T_i$,
and hence the branching rule of the adjoint under the breaking
$SO(8)\to SU(2)^2$ reads
\be
{\bf 28}\to ({\bf 3,1})+({\bf 1,3})+({\bf 2,2})\,,
\ee
corresponding to the breaking
\be
SO(8)\to SO(5)\,.
\ee
It is remarkable that we can obtain a non-regular subgroup of $SO(8)$
by the combination of two {\em inner} automorphisms of $SO(8)$. Each
twist $T_i$ breaks $\mathcal G=SO(8)$ to a regular subgroup $\mathcal
H_i$ (in this case $SU(4)\times U(1)$). However, $T_2$ is not
contained in $\mathcal H_1$, and, although being an inner automorphism
on $\mathcal G$, it acts as an outer automorphisms on $\mathcal H_1$. The
result is the special subgroup $SO(5)$ of $SU(4)\times U(1)$.  All
other gauge groups can, in principle, be calculated along these
lines. For a list of gauge groups that can be obtained this way we
refer the reader to Tab.~6 in Ref.~\cite{Kac:1999gw}.

\section{Breaking $SU(N)$ on the orbifold}
\label{orbifold}

\begin{figure}[htb]
\includegraphics[width=\linewidth]{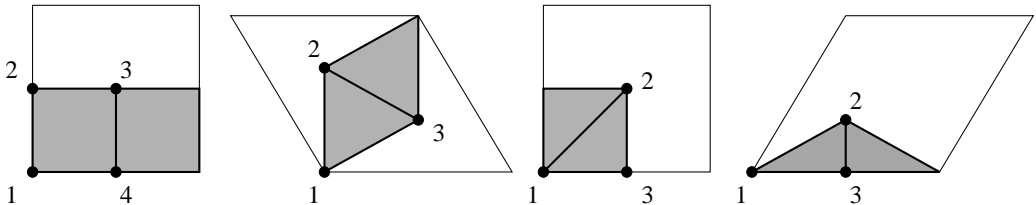}
\caption{\em The four different $Z_n$ orbifold geometries in 6d,
corresponding to $n=2,3,4,6$ (from left to right). We show the
embedding of the orbifold fundamental domain (shaded) in the torus
(thin line) as well as the fixed points (dots). The shaded regions
have to be folded over the center line and the edges (thick lines)
have to be identified. The resulting geometries are ``pillows'' with
three or four corners. Note that the edges correspond to nonsingular
bulk points. }
\label{orbs}
\end{figure}

The torus lattice has a discrete rotational symmetry that can be
modded out to obtain the $T^2/Z_n$ orbifold. The only discrete
rotations possible are of order $n=2,3,4,6$.  The topology of the
resulting spaces are ``pillows'', see Fig.~\ref{orbs}. The two sides
of the pillow represent the bulk and the corners the fixed points. We
depict the four possibilities in Fig.~\ref{orbs}.  Notice that
$T^2/Z_4$ contains two $Z_4$ and one $Z_2$ singularity and $T^2/Z_6$
contains one $Z_2$, $Z_3$ and $Z_6$ singularity each.

In analogy to Eq.~(\ref{torus}), one now introduces orbifold twists
\be 
A_M(p z)=P
A_M(z) P^{-1}\,.
\label{orb}
\ee
where $p$ is the $n$th root of unity
\be
p=\exp(2\pi i/n)\,.
\ee
The additional identification leads to new constraints. Besides the
obvious, $p^n=1$, one also has to take into account that a $Z_n$
rotation followed by a translation along some lattice vector, followed
again by the inverse rotation, equals a lattice translation along the
rotated vector:
\be
p^{-1}(pz+\lambda)=z+p^{-1}\lambda.
\ee
The full set of constraints is thus~\footnote{For the second relation,
notice that the order of the gauge group elements is reversed
w.r.t.~the space group.}
\begin{eqnarray}
T_\lambda T_{\lambda'}&\sim& T_{\lambda'} T_\lambda\sim T_{\lambda+\lambda'}
\,,\label{one}\\
P^{-1} T_\lambda P&\sim& T_{p\lambda}\,,\label{two}\\
P^n&\sim& \one\,.\label{three}
\end{eqnarray}
Here we have introduced the equivalence relation $\sim$ defined as
``equal modulo an element of the center of
$SU(N)$''.\footnote{Mathematically speaking we are looking for special
unitary projective representations of the space group.}  The most
general solution to the first of these constraints has been presented
in the previous section. The main purpose of this paper is to show
that there are nontrivial solutions to the other two constraints,
given torus Wilson lines with generic $m$ and for any
$n=2,3,4,6$.

There is an alternative description to Eq.~(\ref{one}) to
Eq.~(\ref{three}), called the downstairs picture, that only makes
reference to the fundamental domain of the orbifold (i.e.~the physical
space). For any given fixed point $z_f$ of the rotation $p^k$, one can
define a rotation around $z_f$:
\be
p_{z_f}(z)=p^kz+\lambda\,,\qquad \lambda=(1-p^k)z_f
\ee
where $\lambda$ is a lattice vector.  Choosing any fundamental
orbifold domain, the product over the rotations around all four
($Z_2$) or three ($Z_{3,4,6}$) fixed points equals a pure lattice
translation, and some special combinations even yield the trivial one:
choosing the fundamental domains and fixed points labels as in
Fig.~\ref{orbs}, one finds
\begin{align}
p_{z_4}p_{z_3}p_{z_2}p_{z_1}&=1\,,\qquad n=2\,,\label{ds2}\\
p_{z_3}p_{z_2}p_{z_1}&=1\,,\qquad n=3,4,6\label{dsn}\,.
\end{align}
Obviously, any cyclic permutation of these relations hold. For $n=2$,
the anticyclic order also yields one (but not an arbitrary
permutation),\footnote{This can easily be seen by taking the inverse
of Eq.~(\ref{ds2}) and using the fact that $p_{z_f}^2=1$.} while for
$n=3,4,6$ the anticyclic order already yields a nontrivial shift.
Again, these relations must be represented by the corresponding twists:
\begin{align}
P_{z_1}P_{z_2}P_{z_3}P_{z_4}&\sim\one\,,\qquad n=2\,,\nn\\
P_{z_1}P_{z_2}P_{z_3}&\sim \one\,,\qquad n=3,4,6\,,\nn\\
(P_{z_i})^{\nu_i}&\sim \one\,,\label{downstairs}
\end{align}
%
%
with $\nu_i$ being the order of the fixed point $z_i$.  By
re-expressing the lattice shifts through the rotations, it can be
shown that, conversely, the relations Eq.~(\ref{downstairs}) imply
Eqns.~(\ref{one}) to (\ref{three}). In other words, the downstairs
picture (in which we specify the local orbifold twists) is completely
equivalent to the upstairs picture (in which we specify the torus
Wilson lines and the basic $Z_n$ orbifold twist). Moreover, the
downstairs relations can be further reduced by actually solving
Eq.~(\ref{downstairs}) for one of the twists in terms of the
others. In the case of $Z_6$, for instance, the relations then reduce
to
\be
P_{z_2}^3\sim \one\,,\qquad P_{z_3}^2\sim\one\,,
\qquad  (P_{z_2} P_{z_3})^6\sim \one\,.
\ee
While the first two relations are always easy to satisfy, the last
relation becomes highly nontrivial if the two twists do not commute.
In fact, the product $P_{z_3} P_{z_2}$ does not even have to have
finite order. It is possible to generalize the orbifold construction
to allow for gauge twists whose order does not match that of the
spacetime twist\cite{Hebecker:2003jt}. Such models then allow for many
more rank breaking possibilities.  While the downstairs picture is
very useful, in particular for the case of commuting Wilson lines, in
this paper we will mainly stick to the upstairs description. For one,
it makes an important aspect of the new rank breaking mechanism
manifest: it can be viewed as an orbifold of topologically nontrivial
torus Wilson lines. Secondly, the nice factorization of the torus
Wilson lines, obvious from Eq.~(\ref{wilson}), carries over to the
orbifold twists and presents a convenient way to classify 
all possible orbifolds.

The rest of this section is organized as follows. In Sec.~\ref{m0} we
will calculate the $SU(N)$ breaking on the orbifold in the case of
vanishing 't Hooft flux. In particular, we will focus on breakings by
continous Wilson lines, corresponding to the part of the moduli space
of the torus that survives the orbifold projection.  In
Sec.~\ref{generic} we will then show how to construct orbifold twists
that fulfill Eq.~(\ref{two}) and Eq.~(\ref{three}) for generic $m$ and
$N$.

\subsection{The case $m=0$}
\label{m0}

In the case $m=0$, there exists a well defined scheme
\cite{Ibanez:1987xa} to construct solutions to Eq.~(\ref{two}) and
(\ref{three}), by identifying $P$ with a suitable element of the Weyl
group, the symmetry group of the root lattice of the Lie algebra. Such
an element induces an algebra automorphism that maps the Cartan
subalgebra onto itself.
For a given orbifold twist, the Cartan subalgebra naturally decomposes
into two subspaces: The eigenspaces to unit and non-unit eigenvalues
under the linear map $P$. The latter give rise to Wilson lines that
commute with $P$, and Eq.~(\ref{two}) implies that they are discrete.
For $n=2$ one finds
\be T_1^2=T_2^2=\one\,,
\label{discreteWLZ2}
\ee
while for $n=3,4$ one has $T_1=T_2\equiv T$, with
\begin{align}
T^3&=\one\,,\qquad n=3\,,\nn\\
T^2&=\one\,,\qquad n=4\,.
\label{discreteWL}
\end{align}
For $n=6$ there are no discrete Wilson lines.\footnote{In the
downstairs picture this can also be easily understood: any commuting triple
fulfilling Eq.~(\ref{downstairs}) in the case $n=6$ automatically also
satisfies $P_{z_2}=P_{z_1}^2$, $P_{z_3}=P_{z_1}^3$. Hence, the $Z_6$
twist $P_{z_1}$ already determines the other two twists.}  Wilson
lines not invariant under $P$ can still exist and can be constructed
as follows. Consider the shift vector as a map from the torus lattice
to the root lattice, then we can rewrite Eq.~(\ref{two}) as a
composition of maps
\be
P_{adj}^{-1}\circ V=V\circ p\,.
\ee
If the torus lattice can be embedded into the root lattice of the
algebra, one can choose $V$ to be any scalar multiple of that
embedding and identify the rotation $P^{-1}$ with $p$. The Wilson lines
defined this way are thus continous and will break the rank
\cite{Ibanez:1987xa,Forste:2005rs}.  We will not make use of this
description in this paper. Rather, we will consider an equivalent
description in terms of the zero modes of $A_{4,5}$. Just in the case
of the torus, the continous Wilson lines can be transformed into
background VEVs for these extra dimensional components of the gauge
bosons and, hence, parametrize the moduli space of the
compactification. The advantage of this approach is that we can
represent $P$ as a shift (element if the Cartan torus) rather than a
rotation (element of the Weyl group).

Let us consider the case that the gauge twist is the same at each
fixed point (no discrete Wilson lines).  The orbifold shift vector can
be taken, without loss of generality, to be of the form \cite{fuchs}
\be
V=\frac{1}{n} (k_1,\,k_2...k_r)\,,\qquad k_i>0\,,\qquad \sum_i k_i\leq n-1\,.
\label{normal}
\ee
As shown in App.~\ref{moduli}, a flat direction exists if and only if
there are exactly $n-1$ entries with $k_i=1$ with the remaining
$k_i=0$:
\be
V=\frac{1}{n}(0^{r_1},\,1,\,0^{r_2},\,1,\dots 0^{r_{n-1}},\,1,\,0^{r_n})\,.
\label{shiftflat}
\ee
Here, $0^r$ stands for an $r$ dimensional zero vector (some of the
$r_i$ may be zero).  Notice that this means, in particular, that the
inequality in Eq.~(\ref{normal}) is saturated. For $n=2$,
Eq.~(\ref{normal}) already implies a shift vector that is either
trivial or of the form Eq.~(\ref{shiftflat}) and, hence, there are
always flat directions for nontrivial $V$.  For generic $n$, the
breaking pattern induced by this shift vector is
%
\be
SU(N)\to\mathcal H_0\equiv \prod_{i=1}^n SU(N_i)\times U(1)^{n-1}\,,\qquad \sum_{i=1}^n N_i=N\,,
\label{pattern}
\ee
with $N_i=r_i+1$.
%
%
%
%
There are $N_{min}=\min\{N_i\}$ flat directions, which are calculated
in App.~\ref{moduli}. There it is shown that, for vanishing discrete
Wilson lines, a generic point in moduli space breaks $SU(N)$ according
to
\be
SU(N)\to \mathcal H\equiv 
\prod_{i=1}^n U(N_i-N_{min})\times U(1)^{N_{min}-1}\,.
\label{pattern1}
\ee
with 
\be
N\geq n\,,\qquad 
N_k\geq 1\,,\qquad 
N_{min}=\min\{N_k\}\,,\qquad \sum_{k=1}^n {N_k}=N\,.
\ee
The rank of $SU(N)$ is reduced by $N_{min}(n-1)$.

To complete the classification, one could turn on discrete Wilson
lines. The full moduli space of the $T_i=\one$ case survives this
additional projection if and only if the $T_i$ reside in the Cartan
torus of $\mathcal H$. In this case, the unbroken subgroup can be any
full-rank subgroup of $\mathcal H$. It is possible that only a
subspace of the moduli space survives. However, a complete treatment
of these cases lies outside the scope of the present paper and we will
omit it here for brevity. For $Z_6$ there are no discrete Wilson
lines, and our analysis already covers all possible breaking
patterns. The smallest group whose rank can be spontaneously broken in
a $Z_6$ orbifold (with vanishing $m$) is thus $SU(6)$, with a single
modulus breaking all of $SU(6)$.

\subsection{Generic $m$}
\label{generic}

Our classification of solutions to Eqns.~(\ref{two}) and (\ref{three})
for generic $m$ proceeds in two steps. First, we construct the
solution $P_{N,m}$ for $m$, $N$ coprime, which always breaks $SU(N)$ completely,
as we have seen in Sec.~\ref{torons}.  For arbitrary $(N,m)$, we write
the most general solution as
\begin{align}
T_1&=Q_{N/K}\otimes \exp(2\pi i\, W_1)\,,\\
T_2&=(R_{N/K})^{m/K}\otimes \exp(2\pi i\, W_2)\,,\\
P&=P_{N/K,\,m/K}\otimes \exp(2\pi i\, V)\,.
\end{align}
where $W_i$ are discrete Wilson lines subject to
Eqns.~(\ref{discreteWLZ2}) and (\ref{discreteWL}). The shift vectors
$V$ and $W_i$ are elements of the Cartan subalgebra of $SU(K)$.  The
moduli space of this geometry is then given by the moduli space of an
$SU(K)$ theory with vanishing flux. For trivial discrete Wilson lines,
this moduli space has been given in Sec.~\ref{m0} and
App.~\ref{moduli}.

It remains to be shown that, given the Wilson lines~\footnote{We will
drop the indices $N$ and $m$ for the rest of the section.}
\be
T_1=Q,\qquad T_2=R^m\,,
\ee
for $m$ and $N$ coprime, we can actually construct an orbifold
twist $P$ that fulfills Eq.~(\ref{two}) and (\ref{three}).
For $n=2$, it is very easy to write down such a $P$. The matrix
\be
P_{k\ell}=\delta_{k,-\ell}
\label{solz2}
\ee
can easily be confirmed to fullfill the requirements.  
For $n=3,4,6$, we can choose our lattice to be generated by
$\lambda_1=1$ and $\lambda_2=p$.  Relation Eq.~(\ref{two}) then
implies that for any $n=3,4,6$, we must have
\be
P^{-1} Q P \sim R^m
\ee
as well as
\be
P Q P^{-1}\sim\left\{
\begin{array}{lc}
R^{-m}Q^{-1}&n=3\\
R^{-m}&n=4\\
R^{-m}Q&n=6
\end{array}
\right.
\ee
The matrices $Q$ and $R$ have the same eigenvalues, given by the $N$
different $N$th roots of unity. As $m$ and $N$ are coprime, the same
holds true for $R^m$.  As a consequence, one can always find an
$SU(N)$ matrix $U$ that satisfies
\be
U Q U^\dagger \sim R^m.
\label{diag}
\ee
We choose $U$ as~\footnote{This matrix is known as a Vandermonde
matrix. The matrix in Eq.~(\ref{U}) should be divided by its
determinant to obtain an $SU(N)$ matrix, which we ommit here for
clarity.}
\be
U_{k\ell}=N^{-\frac{1}{2}}\, q^{-(k-1)\ell\, m}\,,\qquad q=e^{\frac{2\pi i }{N}}
\label{U}
\ee
The proof that $U$ indeed satisfies Eq.~(\ref{diag}) is presented in
App.~\ref{details}. Moreover, $U$ also satisfies
\be
U^\dagger Q U\sim R^{-m}.
\label{second}
\ee
Notice that $U$ can be multiplied by any diagonal $SU(N)$ matrix from
the left without affecting Eq.~(\ref{diag}), as $R$ is diagonal.
However, Eq.~(\ref{second}) will be modified. 
It can be shown that there is a diagonal $SU(N)$ matrix $X$
satisfying \footnote{We give the precise form of $X$ in
App.~\ref{details}.}
\be
XQX^{\dagger}\sim Q R^m\quad\Leftrightarrow\quad X^\dagger QX \sim Q R^{-m}\,.
\label{X}
\ee
Multiplying $U$ with $X$ we find
\be
( X U)^\dagger  Q (X U) \sim U^\dagger Q R^{-m} U\sim R^{-m}U^\dagger R^{-m}U\sim R^{-m} Q^{-1}
\label{secondz3}
\ee
where in the first step we used Eq.~(\ref{X}), in the second step
Eq.~(\ref{second}) and in the last one Eq.~(\ref{diag}). In a
completely analogous fashion one can show that 
\be
(X^\dagger U)^\dagger Q (X^\dagger U) \sim R^{-m} Q\,.
\label{secondz6}
\ee
One concludes that by choosing 
\be
P^{-1}=\left\{
\begin{array}{lc}
XU&n=3\\
U&n=4\\
X^\dagger U&n=6
\end{array}
\right.
\ee
we satisfy both Eq.~(\ref{diag}) and Eq.~(\ref{second}). It remains
to be shown that 
\be
(XU)^3\sim \one\,,\qquad U^4\sim \one\,,\qquad(X^\dagger U)^6\sim \one\,.
\label{Zn}
\ee
We again postpone the proof of this to App.~\ref{details}. 

Let us illustrate these general considerations with the simplest
possible example: $SU(2)$. The only possible nontrivial choice is
$m=1$. In the adjoint the two Wilson lines read:
%
%
\be
T_1=\diag(+1,-1,-1)\,,\qquad T_2=\diag(-1,-1,+1)\,.\\
\ee
For the $Z_2$ case, Eq.~(\ref{solz2}) actually gives the identity for
$P$. It follows that in this case the local twists are simultaneously
diagonal in the adjoint:
\begin{eqnarray}
P_{z_1}&=&\diag(+1,+1,+1)\,,\nn\\
P_{z_2}&=&\diag(-1,-1,+1)\,,\nn\\
P_{z_3}&=&\diag(-1,+1,-1)\,,\nn\\
P_{z_4}&=&\diag(+1,-1,-1)\,.
\end{eqnarray}
This only happens in the case $N=n=2$. At one fixed point $SU(2)$ is
left unbroken, while at every other fixed point a different $U(1)$
survives. Note that this breaking pattern is qualitatively different
from the usual breaking of $SU(2)$ by continous Wilson lines, as
described in Sec.~\ref{m0}: There the local gauge group is $U(1)$ at
all four fixed points.  For $Z_3$ we find for the twist
\be
P=P_{z_1}=\left(
\begin{array}{ccc}
0&0&1\\
1&0&0\\
0&1&0
\end{array}
\right)\,.
\ee
The local twists are now truly non-commutative as can be seen by
computing the twists associated to the other two fixed points:
\be
PT_1T_2= P_{z_2}=\left(
\begin{array}{ccc}
0&0&-1\\
-1&0&0\\
0&1&0
\end{array}
\right)\,,\qquad
PT_1=P_{z_3}=\left(
\begin{array}{ccc}
0&0&-1\\
1&0&0\\
0&-1&0
\end{array}
\right)\,.
\ee
Geometrically, these twists are $SO(3)$ rotations by $120^o$ around
the axes $(1,1,1)$, $(1,-1,-1)$ and $(-1,-1,1)$ respectively.  Each axis
of rotation defines a $U(1)$ subgroup that remains unbroken at the
corresponding fixed point. It is easy to verify that the product
$P_{z_1}P_{z_2}P_{z_3}$ indeed gives the identity.

In summary, we have seen that the discrete torus Wilson lines that
break $SU(N)$ down to $SU(K)$, with $K$ any divisor of $N$, are
orbifold compatible, i.e.~there exists an orbifold twist that fulfills
Eq.~(\ref{two}) and Eq.~(\ref{three}), for any $n=2,3,4,6$.  In
comparison to the mechanism of rank reduction decribed in Sec.~\ref{m0},
there are no moduli associated to this breaking.  Before concluding this
section we would like to comment on the inclusion of matter to this
scenario. Up to now, we have only considered pure gauge theory or,
more precisely, only fields in the adjoint of the group.  Matter
usually transforms in representations that are sensitive to the center
of the group (such as the fundamental) and, hence, potentially destroy
some or all of the torus configurations.  On the orbifold it is not
uncommon that non-adjoint matter only appears on the fixed points (as,
e.g., in constructions that have extended $\mathcal N=(1,1)$
supersymmetry in 6d). Another possibility is to include other global
or local symmtries in the twists to compensate for the nontrivial
action of the center.

\section{Discussion and Conclusions}
\label{disc}

In this paper we have analyzed orbifolds that break the gauge group
$SU(N)$ to lower rank subgroups. The rank breaking proceeds through
nontrivial toron configurations, meaning the gauge fields have twisted
boundary conditions on the covering torus of the orbifold. These
twisted boundary conditions are of topological nature, characterized
by the 't Hooft flux, and, as a consequence, they cannot be
transformed into a constant background VEV for any extra dimensional
components of the gauge fields. The main result of this paper is that
one can actually orbifold these configurations and that a
classification of all possible breakings emerges from this
approach. Torus Wilson lines can break $SU(N)$ down to $SU(K)$, where
$K$ is a divisor of $N$. The orbifold is compatible with such a
breaking, and the remaining freedom in choosing the orbifold twists is
that of an orbifold with $SU(K)$ gauge group and trivial (commuting)
torus Wilson lines.

As mentioned at the end of Sec.~\ref{torons}, this result can be
generalized almost straightforwardly to the case of other gauge groups
with nontrivial center: the center can be embedded in suitable $SU(N)$
subgroups and the construction of torus and orbifold twists proceeds
as before. They leave an unbroken subgroup that can be orbifolded
in the standard way (i.e., with continous and discrete Wilson lines in
the topologically trivial sector). As a matter of fact, the centers of
groups other than $SU(N)$ are given by abelian groups of order $\leq
4$. Hence, the corresponding twists are particularly simple: they just
correspond to the $SU(N)$ twists desribed in this paper  with
$N\leq 4$.  A more careful treatment of general gauge groups is
postponed to a future publication.

Another possible generalization concerns higher dimensional orbifolds
(based on tori $T^d$ with $d>2$). For $d>2$, the fundamental group of
the adjoint (or, equivalently, the center of the universal cover) is
no longer sufficient to characterize the flat connections on the
torus. In fact, for $SO(N)$ with $N\geq 7$, as well as all expectional
groups, there do exist commuting triples that cannot be simultaneously
conjugated to the same Cartan torus
\cite{Borel:1999bx,Witten:1997bs,Keurentjes:1998uu,Kac:1999gw}. The
surviving unbroken subgroup is therefore rank-reduced. For instance,
the exceptional group $E_8$, which does not have nontrivial pairs,
nevertheless possesses nontrivial triples. It would therefore be
interesting to construct orbifolds based on these nontrivial torus
vacua.\footnote{Note that the asymmetric orbifolds of
Refs.~\cite{deBoer:2001px} are not related to our construction. The
twists employed here can only correspond to symmetric orbifolds in
string theory.}

One can, however, immediately apply our results to 10d orbifolds by
considering particular compactification limits.  Take a heterotic
orbifold with visible gauge group $E_8$. One can think of
compactifying two of the three two-tori, leaving over an effective 6d
theory. It is certainly possible, by making use of standard rank
preserving orbifold breakings, to break $E_8$ to the subgroup
$SO(10)\times SU(4)$ in 6d. In a second step, we break the $SU(4)$
factor completely with our mechanism, while, at the same time, use the
6d orbifold to construct a realistic $SO(10)$ orbifold GUT model. It
is also possible to break to a 6d theory with gauge group $E_6\times
SU(3)$. Standard rank-breaking mechanisms might be used to get the
Standard Model from $E_6$\cite{Forste:2005rs}, while the additional
``flavor'' $SU(3)$ can be broken by the methods described in this
paper.   A more direct application would be an
orbifold reduction of the $SO(32)$ heterotic string to eight
dimension. The $T^2$ compactification has been described in
Ref.~\cite{Lerche:1997rr,Witten:1997bs}, leading to $Sp(16)$ gauge
symmetry in 8d.

Last but not least we would like to comment on an application to
supersymmetry breaking in six dimensions. Minimal $\mathcal N=(1,0)$
supersymmetry has an $R$-symmetry group $SU(2)_R$.  One can break
$SU(2)_R$ and, hence, supersymmetry completely by continous Wilson
lines in the case of $Z_2$ orbifolds but not for $Z_{3,4,6}$. We have
shown that it is nevertheless possible to find discrete Wilson lines
that break all of $SU(2)_R$ for arbitrary $Z_n$, and such a
Scherk-Schwarz mechanism is possible.  Within this context it is
interesting to notice that no continous parameter exists that controls
supersymmetry breaking, yet the breaking is still soft, as locally at
least $N=1$ supersymmetry is preserved at all fixed points.  Similar
constructions can of course be applied to break all or part of
extended supersymmetry.

\section*{Acknowledgments}

I would like to thank M.~Salvatori for useful email exchange. This
work was supported by grants NSF-PHY-0401513, DE-FG02-03ER41271 and
the Leon Madansky Fellowship, as well as the Johns Hopkins Theoretical
Interdisciplinary Physics and Astrophysics Center.

\appendix

\section{Some technicalities}
\label{details}

In this appendix we will prove Eqns.~(\ref{diag}), (\ref{second}),
(\ref{X}), and (\ref{Zn}). Throughout this section $m$ and $N$ are coprime integers and
$q$ is defined as
\be
q=e^{\frac{2 \pi i}{N}}\,.
\ee
Using the property
\be
\sum_{k=1}^{N} q^{\ell k} = N\delta_{\ell,0}\,,\label{polygon}
\ee
Eq.~(\ref{diag}) and Eq.~(\ref{second}) can be readily
verified:
\begin{align}
(U Q U^\dagger)_{k\ell}
&=\frac{q^{-\frac{N-1}{2}}}{N}\sum_{i,j}q^{-(k-1)i\,m+j(\ell-1)\,m}
\delta_{i,j-1}\nn\\
&=\frac{q^{(k-1)m-\frac{N-1}{2}}}{N}\sum_j
\left((q^m)^{\ell-k}\right)^{j}=q^{(k-1)m-\frac{N-1}{2}}\delta_{k\ell}\nn\\
&= q^{\frac{(N-1)(m-1)}{2}} (R^m)_{k\ell}\sim (R^m)_{k\ell}
\end{align}
In the last step of the second line we have made use of the fact that
Eq.~(\ref{polygon}) holds if $q$ is replaced with $q^m$ for $m$ and
$N$ coprime. In the last step we have used that $(N-1)(m-1)$ is always
even.  The proof of Eq.~(\ref{second}) is completely analogous and we
will skip it here.

The identity $U^4\sim U^{\dagger 4}\sim1$ is also quite easy. For $m=1$
\begin{align}
(U^\dagger)^4_{k\ell}&=\frac{1}{N^2}\sum_{i,j,r} q^{k(i-1)+i(j-1)+j(r-1)+r(\ell-1)}\nn\\
&=\frac{1}{N^2}\sum_{i,r} q^{ki -k-i+r\ell-r}\sum_j \left(q^{i+r-1}\right)^j\nn\\
&=\frac{1}{N}\sum_{i,r}q^{ki -k-i+r\ell-r}\delta_{i,1-r}\nn\\
&=\frac{\bar q}{N}\sum_r \left( q^{\ell-k}\right)^r=\bar q \delta_{k\ell}\,.
\end{align}
For $m\neq1$ just replace $q\to q^m$.  Let us now define
\be
X_{k\ell}=
q^{-\frac{k(k-N)}{2}}\, \delta_{k\ell }\,.
\ee
The matrix $X$ does not have unit determinant, $\det X=e^{-\pi i\,
\frac{N^2-1}{3}}$. As in the case of $U$, this can easily be cured
by a rescaling.  Now calculate:
\begin{align}
(XQX^\dagger)_{k\ell}&=q^{-\frac{k(k-N)}{2}-\frac{N-1}{2}+\frac{(k+1)(k-N+1)}{2}}\delta_{k,\ell-1}\nn\\
  &=q^{k-(N-1)}\, \delta_{k,\ell-1}=q\,(RQ)_{k\ell}\sim (RQ)_{k\ell}
\label{XQX}
\end{align}
For $m>1$ one just has to replace $X\to X^m$, which concludes our proof of Eq.~(\ref{X}).

To prove the remaining relations in Eq.~(\ref{Zn}) we will need the
identity\footnote{The fact that $\bar Z Z=N$ can be inferred by
considering the Discrete Fourier Transformation (DFT) of
$x_k=q^{(k-N/2)^2/2}$. By performing the DFT and its inverse, one
finds $x_k=\bar ZZ/N \, x_k$. We shall not prove the value of the
phase in Eq.~(\ref{z}) since it will turn out to be irrelevant (see comment
after Eq.~(\ref{last})).}
\be
Z\equiv\sum_{k=0}^{N-1} q^{\frac{(k-N/2)^2}{2}}=\sqrt{iN}
\label{z}
\ee
Let us start with $m=1$. 
\begin{align}
(XU)_{k\ell}^3&=N^{-3/2}\sum_{i,j} q^{-\frac{k(k-N)}{2}-(k-1)i-\frac{i(i-N)}{2}-(i-1)j
-\frac{j(j-N)}{2}-(j-1)\ell}\nn\\
&=N^{-3/2}\sum_{i,j} q^{-\frac{(i+j+k-1-N/2)^2}{2}+\frac{(j+k-1-N/2)^2}{2}-\frac{k(k-N)}{2}+j-\frac{j(j-N)}{2}-(j-1)\ell}\nn\\
  &=N^{-1}(i)^{-\frac{1}{2}}
\sum_j q^{(j-1)(l-k)+\frac{(1+N/2)^2}{2}}=(i)^{-\frac{1}{2}}q^{\frac{(1+N/2)^2}{2}}\delta_{k,\ell}\,.
\label{last}
\end{align}
The fact that we have collected a nontrivial phase (i.e. not an
integer power of $q$) is related to the fact that our matrices $X$ and
$U$ are $U(N)$ as opposed to $SU(N)$ matrices. This could easily
remedied by a rescaling, without affecting the other relations
Eq.~(\ref{diag}), (\ref{second}), and (\ref{X}). Since $SU(N)$ is a
group, it follows that the r.h.s.~of Eq.~(\ref{last}) has to be an
$SU(N)$ element also. After a suitable rescaling we thus arrive at the
first relation in Eq.~(\ref{Zn}). For the last relation in
Eq.~(\ref{Zn}) we calculate
\begin{align}
(X^\dagger U)_{k\ell}^3&
=N^{-3/2}\sum_{i,j} q^{\frac{k(k-N)}{2}-(k-1)i+\frac{i(i-N)}{2}-(i-1)j
+\frac{j(j-N)}{2}-(j-1)\ell}\nn\\
&=N^{-3/2}\sum_{i,j} q^{\frac{(i-j-k+1-N/2)^2}{2}-\frac{(j+k-1+N/2)^2}{2}+\frac{k(k-N)}{2}+j+\frac{j(j-N)}{2}-(j-1)\ell}\nn\\
  &=N^{-1}(i)^{\frac{1}{2}}
\sum_j q^{-j (-2 + k + l + N)+k+l-kN-\frac{(1-N/2)^2}{2}}\nn\\
&=(i)^{\frac{1}{2}}q^{2-Nk-\frac{(1-N/2)^2}{2}}\delta_{k,2-\ell}\,.
\end{align}
For $N=2$, this is already proportional to the identity. For $N>2$ we
square this to find
\be
(X^\dagger U)_{k\ell}^6=iq^{4-(1+N/2)^2}\delta_{k,\ell}\,.
\ee
Finally, for $m>1$ we can just replace $q\to q^m$ and observe that Eq.~(\ref{z}) still holds since $m$ and $N$ are coprime.

\section{The moduli space for m=0}

\label{moduli}

In this appendix we would like to calculate the moduli space on the
orbifold, in the case $m=0$.  To this end, we calculate the scalar
zero modes from the projection Eq.~(\ref{shiftflat}) and subsequently
find those modes that correspond to flat directions in the potential.
The potential is coming from the term
\be
\mathcal V\sim \Tr (F_{ij}F^{ij})=2 g^{-1} 
\Tr F_{45}^2=-2 g^{-1} \Tr [A_4,A_5]^2=4 g^{-1}\Tr [A_+,A_-]^2\,,
\ee
where $g=\det g_{ij}$ and we have defined the complex scalars
$A_\pm=A_4\pm i A_5$. Notice that the hermiticity of the $A_i$ implies
the reality constraint $A_+^{\ \dagger}=A_-$. The orbifold boundary
conditions now read:
\be
A_\pm(pz)=\exp\left(2 \pi i \left[V\mp\frac{1}{n}\right]\right) A_\pm(z)\,.
\label{scalars}
\ee
The zero modes correspond to those states where the term in the square
brackets in Eq.~(\ref{scalars}) is integer.

To find these zero modes, note that there are $n$ special roots that
have $V\cdot\alpha=1/n\mod \mathbb Z$: the $n-1$ simple roots that
have $k_i=1$ in Eq.~(\ref{shiftflat}), as well as the most negative
root (defined as minus the sum of all simple roots). They all belong
to different irreducible representations of the subgroup $\mathcal
H_0$ defined in Eq.~(\ref{pattern}). By inspection of the remaining
roots and their associated raising and lowering operators,\footnote{
The positive roots of $SU(N)$ are given by $\{\alpha_{\ell
k}=\alpha_{(\ell)}+\alpha_{(\ell+1)}+\dots+\alpha_{(k)}\,,\
0\leq\ell\leq k\leq r\}$ in terms of the simple roots
$\alpha_{(i)}$. The associated creation operator is given by
$(E^+_{\ell k})_{ij}=\delta_{i,\ell}\delta_{j,k+1}$.} one can
parametrize the zero modes of $A_+$ as
\be
A_+=
\left(
\begin{array}{cccccc}
0&A_1&0  &0&\cdots&0\\
0&0  &A_2&0&\cdots&0\\
\vdots&&&&&\vdots\\
0&0&0&0&\cdots&A_{n-1}\\
A_n&0&0&0&\cdots&0
\end{array}
\right)\,.
\ee
Here the entry in the $i$th row and $j$th column is a matrix of
dimension $N_i\times N_j$. In particular, the $A_i$ are $N_i\times
N_{i+1}$ matrices forming the representation
\be
A_i=({ N_i},{\bar N_{i+1}})\,,
\ee
where we have adopted a cyclic convention for the indices. One
immediately calculates $F_{+-}=[A_+,A_-]$
\be
F_{+-}=
\left(
\begin{array}{cccc}
A_1\bar A_1-\bar A_n A_n\\
&A_2\bar A_2-\bar A_1 A_1\\
&&\ddots\\
&&&A_n\bar A_n-\bar A_{n-1} A_{n-1}
\end{array}
\right)\,.
\label{F}
\ee
The diagonal blocks are now square matrices of dimension $N_i$.  The
vanishing of $F_{+-}$ is a neccesary and sufficient condition for the
potential 
\be
\mathcal V\sim \sum_{i=1}^n \Tr (A_i\bar A_i-\bar A_{i-1}A_{i-1})^2
\label{V}
\ee
to possess a flat direction.  One can always use the $\mathcal H_0$
gauge symmetry to diagonalize all $A_i\bar A_i$. If there is a flat
direction, then in this basis the matrices $\bar A_i A_i$ must be
diagonal as well.  Let us define $N_{min}=\min\{N_i\}$. Then all
matrices $A_i \bar A_i$ and $\bar A_i A_i$ have at least rank
$N_{min}$. One concludes that if a flat direction exists, without loss
of generality one can assume:
\be
A_i \bar A_i=\bar A_{i-1} A_{i-1}=
\operatorname{diag}(a_1,\dots,a_{N_{min}},0^{N_i-N_{min}})\,,\\
\label{flat}
\ee
where the $a_k$ are real constants.  All that remains to show is that
there exists a configuration $A_i$ that fulfills
Eq.~(\ref{flat}). This can easily be achieved by choosing the first
$N_i$ diagonal entries of $A_i$ equal to $\sqrt a_i$ with all other
entries equal to zero. A generic VEV along this flat direction breaks
each $SU(N_k)$ factor to $SU(N_k-N_{min})$. To obtain the $U(1)$
factors, it is sufficient to find the rank of the surviving
subgroup, i.e., we are looking for the number of Cartan generators that satisfy
\be
[A_+,H]=0\,.
\ee
To this end, note that we can view the quantity $A_+$ as a linear map
from the Cartan subalgebra to the subspace of $\mathfrak{su} (N)$
generated by those $E_\alpha$ that are nonzero in $A_+$. Writing down
the matrix corresponding to that map, it can be read off that it has
rank $N_{min}(n-1)$. The rank-nullity theorem then states that the
dimension of the kernel of that map is equal to $N-1-N_{min}(n-1)$,
which must equal the rank of the surviving subgroup. Thus, the
breaking pattern turns out to be
\be
SU(N)\to 
\prod_{k=1}^n SU(N_k)\times U(1)^{n-1}
\to \prod_{k=1}^n U(N_k-N_{min})\times U(1)^{N_{min}-1}\,.
\label{pattern2}
\ee
It may be verified that the rank of this group is indeed
$N-1-N_{min}(n-1)$.  Let us summarize the conditions the different
quantities in Eq.~(\ref{pattern2}) are subject to:
\be
N\geq n\,,\qquad 
N_k\geq 1\,,\qquad 
N_{min}=\min\{N_k\}\,,\qquad \sum_{k=1}^n {N_k}=N\,.
\ee

Let us now turn to shift vectors that are not of the form
Eq.~(\ref{shiftflat}) but still fulfill condition (\ref{normal}). The
breaking pattern will still be of the form Eq.~(\ref{pattern}), but
now with fewer $SU(N_k)$ factors. The simple roots and the most
negative root still belong to bifundamentals. The important difference
is that one or more of these bifundamentals cease to have zero modes
(some $k_i>1$ and/or $\sum k_i<n-1$). Removing one or more of the
$A_i$ from Eq.~(\ref{F}) or (\ref{V}) clearly destroys the possibility
of having flat directions.  We conclude that flat directions exist if
and only if $V$ is equivalent to the form Eq.~(\ref{shiftflat}).

\end{document}